\documentclass[10pt]{revtex4}
\newcount\driver
\newcount\bozza

\font\ottorm=cmr8\font\ottoi=cmmi8\font\ottosy=cmsy8%
\font\ottocss=cmcsc8%
\font\sixrm=cmr6\font\sixi=cmmi6\font\sixsy=cmsy6%
\font\fiverm=cmr5\font\fivesy=cmsy5
\font\fivei=cmmi5
\font\tenmib=cmmib10
\font\sevenmib=cmmib10 scaled 800

 2

\font\cs=cmcsc10
\font\sc=cmcsc10

\font\elevenrm=cmr11
\font\twelverm=cmr12
\font\ottorm=cmr8
\font\msytw=msbm9 scaled\magstep1

\font\indbf=cmbx10 scaled\magstep2

\font\ottorm=cmr8\font\ottoi=cmmi8\font\ottosy=cmsy8%
\font\ottocss=cmcsc8%
\font\sixrm=cmr6\font\sixi=cmmi6\font\sixsy=cmsy6%
\font\fiverm=cmr5\font\fivesy=cmsy5
\font\fivei=cmmi5

\def\ottopunti{\def\rm{\fam0\ottorm}%
\textfont0=\ottorm\scriptfont0=\sixrm\scriptscriptfont0=\fiverm%
\textfont1=\ottoi\scriptfont1=\sixi\scriptscriptfont1=\fivei%
\textfont2=\ottosy\scriptfont2=\sixsy\scriptscriptfont2=\fivesy%
\textfont4=\ottocss\scriptfont4=\sc\scriptscriptfont4=\sc%
\scriptfont4=\ottocss\scriptscriptfont4=\ottocss%
\textfont5=\tenmib\scriptfont5=\sevenmib\scriptscriptfont5=\fivei
\setbox\strutbox=\hbox{\vrule height7pt depth2pt width0pt}%
\normalbaselineskip=9pt\let\sc=\sixrm\normalbaselines\rm}

\mathchardef\BDpr = "0540  %Dpr
\mathchardef\Bg   = "050D  %gamma

{\count255=\time\divide\count255 by 60 \xdef\hourmin{\number\count255}
        \multiply\count255 by-60\advance\count255 by\time
   \xdef\hourmin{\hourmin:\ifnum\count255<10 0\fi\the\count255}}

\def\openone{\leavevmode\hbox{\elevenrm 1\kern-3.63pt\twelverm1}}%
\def\*{\vglue0.5truecm}

\let\a=\alpha \let\b=\beta  \let\g=\gamma  \let\d=\delta \let\e=\varepsilon
      \let\k=\kappa 
        \let\x=\xi     \let\p=\pi    \let\r=\rho
    \let\f=\varphi 
   \let\o=\omega
 \let\D=\Delta  \let\L=\Lambda 
         
\let\O=\Omega 

\def\\{\hfill\break} \let\==\equiv

\let\io=\infty 

\let\0=\noindent

\def\media#1{{\langle#1\rangle}}
\def\ie{{i.e. }}
\let\dpr=\partial

\def\tende#1{\,\vtop{\ialign{##\crcr\rightarrowfill\crcr
 \noalign{\kern-1pt\nointerlineskip} \hskip3.pt${\scriptstyle
 #1}$\hskip3.pt\crcr}}\,}
\def\circage{\lower2pt\hbox{$\,\buildrel > \over {\scriptstyle \sim}\,$}}
\def\otto{\,{\kern-1.truept\leftarrow\kern-5.truept\to\kern-1.truept}\,}

\def\T#1{{#1_{\kern-3pt\lower7pt\hbox{$\widetilde{}$}}\kern3pt}}
\def\VVV#1{{\underline #1}_{\kern-3pt
\lower7pt\hbox{$\widetilde{}$}}\kern3pt\,}
\def\W#1{#1_{\kern-3pt\lower7.5pt\hbox{$\widetilde{}$}}\kern2pt\,}

\def\indica{\leaders \hbox to 0.5cm{\hss.\hss}\hfill}
\def\guida{\leaders\hbox to 1em{\hss.\hss}\hfill}

\def\hhh{{\bf h}}

\def\defin{{\buildrel def\over=}}

\mathchardef\aa   = "050B
\mathchardef\bb   = "050C
\mathchardef\ggg  = "050D
\mathchardef\xxx  = "0518
\mathchardef\zzzzz= "0510
\mathchardef\oo   = "0521
\mathchardef\lll  = "0515
\mathchardef\mm   = "0516
\mathchardef\Dp   = "0540
\mathchardef\H    = "0548
\mathchardef\FFF  = "0546
\mathchardef\ppp  = "0570
\mathchardef\Bn   = "0517
\mathchardef\pps  = "0520
\mathchardef\fff  = "0527
\mathchardef\FFF  = "0508
\mathchardef\nnnnn= "056E

\def\to{\rightarrow}

\def\qed{\hfill\raise1pt\hbox{\vrule height5pt width5pt depth0pt}}

\def\indic{\hbox{\raise-2pt \hbox{\indbf 1}}}

\def\RRR{\hbox{\msytw R}}

 \def\ZZZ{\hbox{\msytw Z}}

\def\V0{{\bf 0}}

\font\tenmib=cmmib10 
\font\sevenmib=cmmib7\font\fivemib=cmmib5 

\font\fivei=cmmi5\font\sixi=cmmi6\font\ottoi=cmmi8
\font\ottorm=cmr8\font\fiverm=cmr5\font\sixrm=cmr6
\font\ottosy=cmsy8\font\sixsy=cmsy6\font\fivesy=cmsy5%%
\font\ottocss=cmcsc8%

\mathchardef\Ba   = "050B  %alfa
\mathchardef\Bb   = "050C  %beta
\mathchardef\Bg   = "050D  %gamma
\mathchardef\Bd   = "050E  %delta
\mathchardef\Be   = "0522  %varepsilon
\mathchardef\Bee  = "050F  %epsilon
\mathchardef\Bz   = "0510  %zeta
\mathchardef\Bh   = "0511  %eta
\mathchardef\Bthh = "0512  %teta
\mathchardef\Bth  = "0523  %varteta
\mathchardef\Bi   = "0513  %iota
\mathchardef\Bk   = "0514  %kappa
\mathchardef\Bl   = "0515  %lambda
\mathchardef\Bm   = "0516  %mu
\mathchardef\Bn   = "0517  %nu
\mathchardef\Bx   = "0518  %xi
\mathchardef\Bom  = "0530  %omi
\mathchardef\Bp   = "0519  %pi
\mathchardef\Br   = "0525  %ro
\mathchardef\Bro  = "051A  %varrho
\mathchardef\Bs   = "051B  %sigma
\mathchardef\Bsi  = "0526  %varsigma
\mathchardef\Bt   = "051C  %tau
\mathchardef\Bu   = "051D  %upsilon
\mathchardef\Bf   = "0527  %phi
\mathchardef\Bff  = "051E  %varphi
\mathchardef\Bch  = "051F  %chi
\mathchardef\Bps  = "0520  %psi
\mathchardef\Bo   = "0521  %omega
\mathchardef\Bome = "0524  %varomega
\mathchardef\BG   = "0500  %Gamma
\mathchardef\BD   = "0501  %Delta
\mathchardef\BTh  = "0502  %Theta
\mathchardef\BL   = "0503  %Lambda
\mathchardef\BX   = "0504  %Xi
\mathchardef\BP   = "0505  %Pi
\mathchardef\BS   = "0506  %Sigma
\mathchardef\BU   = "0507  %Upsilon
\mathchardef\BF   = "0508  %Fi
\mathchardef\BPs  = "0509  %Psi
\mathchardef\BO   = "050A  %Omega
\mathchardef\BDpr = "0540  %Dpr
\mathchardef\Bstl = "053F  %*

\def\V#1{{\bf#1}}
\let\aa=\Ba\let\fff=\Bf\let\defin=\defi
\let\oo=\Bo\let\nn=\Bn
\let\pps=\Bps\def\hhh={\V h}
\let\bb=\Bb

\def\RRR{\hbox{\msytw R}}

 \def\ZZZ{\hbox{\msytw Z}}

\def\Tr{{\rm Tr}}
\def\dist{{\rm dist}}

\def\ins#1#2#3{\vbox to0pt{\kern-#2 \hbox{\kern#1 #3}\vss}\nointerlineskip}
\newdimen\xshift \newdimen\xwidth \newdimen\yshift
\newcount\griglia

\def\insertplot#1#2#3#4#5#6{%
\begin{figure}[h]
\begin{center}
\vspace{#2pt}
\begin{minipage}{#1pt}
#3
\ifnum\driver=1
\griglia=#6
\ifnum\griglia=1
\openout13=griglia.ps
\write13{gsave .2 setlinewidth}
\write13{0 10 #1 {dup 0 moveto #2 lineto } for}
\write13{0 10 #2 {dup 0 exch moveto #1 exch lineto } for}
\write13{stroke}
\write13{.5 setlinewidth}
\write13{0 50 #1 {dup 0 moveto #2 lineto } for}
\write13{0 50 #2 {dup 0 exch moveto #1 exch lineto } for}
\write13{stroke grestore}
\closeout13
\includegraphics{griglia.ps}\fi
\includegraphics{#4.ps}\fi
\ifnum\driver=2
\fi
\end{minipage}
\end{center}
\caption{#5}
\end{figure}
}

\newdimen\shift \shift=-1truecm
\def\lb#1{%
\ifnum\bozza=1
\label{#1}\rlap{\kern\shift{$\scriptstyle#1$}}
\else\label{#1}
\fi}

\def\be{\begin{equation}}
\def\ee{\end{equation}}
\def\bea{\begin{eqnarray}}\def\eea{\end{eqnarray}}
\def\bean{\begin{eqnarray*}}\def\eean{\end{eqnarray*}}
\def\bfr{\begin{flushright}}\def\efr{\end{flushright}}
\def\bc{\begin{center}}\def\ec{\end{center}}
\def\ba#1{\begin{array}{#1}} \def\ea{\end{array}}
\def\bd{\begin{description}}\def\ed{\end{description}}
\def\bv{\begin{verbatim}}\def\ev{\end{verbatim}}
\def\nn{\nonumber}
\def\Halmos{\hfill\vrule height10pt width4pt depth2pt \par\hbox to \hsize{}}

\def\const{{\rm const\,}}

\renewcommand{\theequation}{\arabic{section}.\arabic{equation}}

\newdimen\xshift \newdimen\xwidth \newdimen\yshift \newdimen\ywidth

\def\ins#1#2#3{\vbox to0pt{\kern-#2\hbox{\kern#1 #3}\vss}\nointerlineskip}

\def\eqfig#1#2#3#4#5{
\par\xwidth=#1 \xshift=\hsize \advance\xshift
by-\xwidth \divide\xshift by 2
\yshift=#2 \divide\yshift by 2
\line{\hglue\xshift \vbox to #2{\vfil
#3 \includegraphics{#4.ps}
}\hfill\raise\yshift\hbox{#5}}}

\def\8{\write12}

\begin{document}

\title{Ground state energy of the low density Hubbard model.\\
An upper bound.}

\*

\author{Alessandro Giuliani}
\affiliation{Department of Physics, 
Princeton University, Princeton 08544 NJ, USA}
\vspace{1cm}
\date{\today}

\begin{abstract} 
We derive an upper bound on the ground state energy of the three-dimensional
(3D) repulsive Hubbard model on the cubic lattice agreeing in the low 
density limit with the known asymptotic expression of the ground state energy 
of the dilute Fermi gas in the continuum. As a corollary, we prove an old 
conjecture on the low density behavior of the 3D Hubbard model,
i.e., that the total spin of the ground state vanishes as the density goes 
to zero.
\end{abstract}

\maketitle

%%%%%%%%%%%%%%%%%%%%%%%%%%%%%%%%%%%%%%%%%%%%%%%%%%%%%%%%%%%%%%%%%%%%%%%%%%
%%%%%%%%%%%%%%%%%%%%%%%%%%%%%%%%%%%%%%%%%%%%%%%%%%%%%%%%%%%%%%%%%%%%%%%%%%
\section{Introduction}
%%%%%%%%%%%%%%%%%%%%%%%%%%%%%%%%%%%%%%%%%%%%%%%%%%%%%%%%%%%%%%%%%%%%%%%%%%
%%%%%%%%%%%%%%%%%%%%%%%%%%%%%%%%%%%%%%%%%%%%%%%%%%%%%%%%%%%%%%%%%%%%%%%%%%

Recent developments in the theory of low density Bose and Fermi gases
made it possible to verify old conjectures on the leading asymptotics for the 
ground state energy of dilute gases of bosons or fermions in the continuum, 
interacting with positive short range potentials. While the heuristing argument
suggesting that the ground state energy of the 3D hard-core Bose gas is
proportional to the scattering length $a$ of the potential goes back to Lenz
[Le], the first ideas 
in the direction of proving that Lenz's formula is correct in the low density
limit are due to Dyson [D], who first established an asymptotically correct 
upper bound and a rigorous (but 14 times too small)
lower bound for the hard core Bose gas in 3 dimensions. An asymptotically 
correct lower bound was proven much more recently by 
Lieb and Yngvason [LY]. Their work inspired much of the recent developments
in the rigorous theory of low density quantum many body systems, 
see [LSSY] for a comprehensive review of the subject till 2005.
In particular, a result that we would like to mention,
strictly related to the problem studied in this paper,
is the proof in [LSS] that the ground state energy per unit volume
of the 3D Fermi gas in the continuum with short 
range repulsive interaction (and scattering length $a>0$) is given, 
in the low density limit $\r a^3\to 0$, by:
\be e(\r_{\uparrow},\r_{\downarrow})=\frac{\hbar^2}{2m}\frac35(6\p^2)^{2/3}
(\r_{\uparrow}^{5/3}+\r_{\downarrow}^{5/3})+\frac{\hbar^2}{2m}8\p a\r_{
\uparrow}\r_{\downarrow}+o(a\r^2)\label{0.1}\ee
where $\r_{\uparrow,\downarrow}$ are the densities of spin up and spin down 
particles and $m$ is their mass. Moreover $\r=\r_{\uparrow}+\r_{\downarrow}$
and $o(a\r^2)$ is a suitable function of the total 
density $\r$ and of the scattering length $a$ vanishing faster than $a\r^2$
in the limit $\r a^3\to 0$. 

It is very natural to ask whether a formula similar to (\ref{0.1}) is valid 
for a dilute Fermi gas with short range repulsive interaction on the lattice
and, in particular, for the most popular model for correlated electrons
in condensed matter physics: the Hubbard model. The Hubbard model
is the simplest possible lattice model of interacting electrons
displaying many ``real world'' features and in the last 40 years it has
been subject of intense research efforts. Nonetheless, even its qualitative 
behavior in 2 or 3 dimensions is far from clear and there are very few
rigorous results available in the literature. A survey of known results and 
open problems in the Hubbard model can be found in [Li] and in [T]. 

In the present paper we shall
derive an upper bound for the ground state energy of the 3D repulsive Hubbard
model on the cubic lattice 
with the same asymptotic behavior as (\ref{0.1}). As a corollary we 
shall prove one of the open problems posed by Elliott Lieb in his review 
article on the Hubbard model (see [Li, Problem 3]). More precisely, 
we shall prove the following old conjecture on the low density behavior of 
the 3D Hubbard model.\\

{\bf Proposition.} {\it 
Let $S_{max}=N_{tot}/2$ be the maximum spin a system of $N_{tot}$
electrons can achieve and let $S$ be the spin of the
ground state of the 3D repulsive Hubbard model in a cubic box 
$\L\subset\ZZZ^3$ in presence of $N_{tot}$ electrons 
(or the maximum such spin in case of degeneracy). Then
\be\lim_{\r\to 0}\lim_{|\L|\to\io}S/S_{max}=0\,,\label{oldconjecture}\ee
where the thermodynamic limit is taken keeping the total density $\r=N_{tot}/
|\L|$ fixed.}\\

{\bf Remark.} A sketch of the proof of this claim
already appeared in [BLT]: in this note we provide all the details of 
the necessary computations as well as an explicit bound on the rate of 
convergence (see Corollary 1 below). Our strategy imitates the one in [LSS].\\

The paper is organized as follows. In the next two subsections we shall 
introduce the model, introduce the notion of scattering length and state the 
main results, i.e., the upper bound on the ground state energy and an explicit 
bound on the rate of convergence of $S/S_{max}$ to $0$ in 
(\ref{oldconjecture}). In Sec.II and in the two Appendices we shall give the 
proof.

%%%%%%%%%%%%%%%%%%%%%%%%%%%%%%%%%%%%%%%%%%%%%%%%%%%%%%%%%%%%%%%%%%%%%%%%%%
%%%%%%%%%%%%%%%%%%%%%%%%%%%%%%%%%%%%%%%%%%%%%%%%%%%%%%%%%%%%%%%%%%%%%%%%%%
\subsection{The model}
%%%%%%%%%%%%%%%%%%%%%%%%%%%%%%%%%%%%%%%%%%%%%%%%%%%%%%%%%%%%%%%%%%%%%%%%%%
%%%%%%%%%%%%%%%%%%%%%%%%%%%%%%%%%%%%%%%%%%%%%%%%%%%%%%%%%%%%%%%%%%%%%%%%%%

Given a cubic lattice $\L$ of lattice spacing $r_0$, the Hamiltonian 
of the Hubbard model on $\L$ 
for $N$ spin-up particles and $M$ spin-down particles can be written as:
\be H=-\D_X-\D_Y+U v_{XY}\label{hamiltonian}\ee
where:\\
1) $X=(x_1,\ldots,x_{N})$ and $Y=(y_1,\ldots,y_M)$ are the coordinates
of the spin-up and spin-down particles, respectively;\\
2) $\D_X=\sum_{i=1}^{N}\D_{x_i}$, $\D_Y=\sum_{j=1}^{M}\D_{y_j}$ and
$\D_xf(x)=r_0^{-2}\sum_{x':|x'-x|=r_0}(f(x')-f(x))$;\\
3) $v_{XY}=\sum_{i=1}^{N}\sum_{j=1}^{M}\d_{x_i,y_j}$;\\
4) $U\ge 0$;\\
5) $H$ acts on the space of functions antisymmetric in the $X$ and in the $Y$
coordinates separately and vanishing outside the box $\L$ (Dirichlet boundary
conditions).\\

{\bf Remark.}
In this note we restrict for simplicity to the case of a nearest neighbor 
hopping and a delta interaction, however the analysis below can be generalized 
to cases with different hopping terms and different short range interactions 
(not necessarily zero - or finite - range).\\

We want to obtain an upper bound for the ground state energy
that is asymptotically correct, at the lowest order,
as $\r a^3\to 0$, where $\r=(N+M)/|\L|$ and $a$ is the scattering
length of the potential. The latter can be conveniently defined in terms 
of the solution to the zero energy scattering equation 
\be -\D_x\f(x)+\frac{U}2\d_{x,0}\f(x)=0\ee
subject to the boundary condition $\lim_{|x|\to\io}\f(x)=1$. The solution is
\be \f(x)=1-4\p \frac{a}{r_0}\int_{|k_i|\le\p r_0^{-1}}
\frac{d^3k}{(2\p r_0^{-1})^{3}}\ 
\frac{e^{ikx}}{2\sum_{i=1}^3(1-\cos k_ir_0)}\label{scattering}\ee 
where the coefficient $a$ has the interpretation of scattering length and is 
given by 
\be 8\p a=r_0\ \frac{Ur_0^2}{Ur_0^2\g+1}\;,\qquad
\g=\frac12\int_{|k_i|\le\p}\frac{d^3k}{(2\p)^{3}}\ 
\frac{1}{2\sum_{i=1}^3(1-\cos k_i)}\;.\label{length}\ee
Note that $\lim_{|x|\to\io}(1-\f(x))|x|=a$ (this means that at large distances
$\f(x)$ looks very much like the scattering solution in the continuum, \ie
$\f(x)\simeq1-a/|x|$ at large distances) and that 
$8\p a\le U r_0^3$ (this is the analogue of the inequality
of Spruch and Rosenberg [SR] in the lattice case). Another important remark
is that, given 
any simply connected domain $\O$ containing the origin, the ``flux''
of the discrete derivative of $\f(x)$ across the boundary of $\O$ is 
independent of $\O$ and equal to $4\p a$:
\be \sum_{<x,x'>}^{(\dpr\O)}(\f(x')-\f(x))=4\p a\label{0.2}\ee
where $\sum_{<x,x'>}^{(\dpr\O)}$ is the sum over the bonds connecting nearest
neighbor sites with $x\in\O$ and $x'\in\O^c$. This simply follows by 
the remark that $\D\f(x)=0$, $\forall x\neq0$, and by discrete ``integration
by parts''. 

\subsection{Main results}

We are now ready to state our main result.\\

{\bf Theorem 1.} {\it Fix $\r_{\uparrow}  =N/|\L|$, $\r_\downarrow=M/|\L|$ and
$\rho = \r_{\uparrow}+\r_\downarrow$,
and let $E_0(N,M,\L)$ denote the ground state energy of $H$
with the appropriate antisymmetry in each of the $N,M$ coordinate variables.
Then, for small $\rho a^3$,
\be\lim_{|\L|\to \infty} \frac{ 1}{|\L|} E_0(N,M,\L)\le e_0(\r_{\uparrow},\r_{
\downarrow})+8\pi  a 
\r_{\uparrow}\r_{\downarrow}
+a\rho^2 \e(\rho a^3),\label{0.3}\ee
where $e_0(\r_{\uparrow},\r_{\downarrow})$ is the specific ground state energy
of the free Fermi gas on the lattice, i.e., of (\ref{hamiltonian})
with $U=0$, and $0\le \e(\rho a^3)\le \const(\r^{1/3}a)^{2/9}$.}\\

{\bf Remarks.}\\
1) The theorem is valid for any repulsion strength $U\ge 0$, 
including the limiting case $U=+\io$ of infinite repulsion.\\
2) If $\r r_0^3\ll1$, the specific ground state energy
of the free Fermi gas on the lattice can be written as 
$$e_0(\r_{\uparrow},\r_{\downarrow})=\frac 35 \left( 6\pi^2 \right)^{2/3} 
(\r_{\uparrow}^{5/3}+\r_{\downarrow}^{5/3})+\const\, r_0^2\r^{7/3}$$
Then, as long as $a/r_0\gg\sqrt{\r^{1/3}a}$, in the r.h.s. of (\ref{0.3}) 
we can replace $e_0(\r_{\uparrow},\r_{
\downarrow})$ by $\frac 35 \left( 6\pi^2 \right)^{2/3} 
(\r_{\uparrow}^{5/3}+\r_{\downarrow}^{5/3})$ and still have an error term that
is much smaller than $\r^2 a$. In this case the upper bound (\ref{0.3})
looks precisely the same as (\ref{0.1}). \\
3) It would be nice to establish that the r.h.s. of (\ref{0.3}) is 
the correct low density behavior of the ground state energy of the 3D Hubbard 
model. In order to prove this we should provide a lower bound with the same 
asymptotic behavior as the r.h.s. of (\ref{0.3}). The natural idea would be 
to proceed as in the continuum case [LSS], that is by 
exploiting Dyson's idea of replacing the ``hard'' interaction potential by 
a ``soft'' one, at the expense of using up some kinetic energy. Of course, 
in order to get the correct 0--th order contribution in the lower bound, 
we need to use at least part of the kinetic energy to ``fill the Fermi sea'':
so technically one of the main steps in the proof of the lower bound in 
[LSS] is the proof of a ``Dyson Lemma'' in presence of an infrared cutoff, 
allowing for a replacement of the hard interaction by a soft one, at the 
expense only of the high momentum part of the kinetic energy. We would expect 
that this result is actually independent of the presence or absence of
an underlying lattice structure: however the proof of the ``Dyson Lemma'' with 
infrared cutoff in [LSS] uses in a crucial way rotational invariance of the 
problem and it is an open problem to adapt it to the lattice case.\\

The result of the Theorem above, 
combined with the remark that the first term in the r.h.s. of (\ref{0.3})
provides an obvious lower bound to the ground state energy, implies that 
$|E_0(N,M,\L)-E^{(U=0)}_0(N,M,\L)|\le 
\const |\L|a\r^2$ and this
in turns implies that, if we fix the total density $\r$ and minimize the 
energy over the possible choices of $\r_{\uparrow,\downarrow}$, we find that 
at low density the absolute ground state satisfies $|\r_{\uparrow}-
\r_{\downarrow}|\le \const \r \sqrt{\r^{1/3}a}$
This implies that the total spin $S$
of the ground state satisfies the following.\\

{\bf Corollary 1.} {\it Let $S$ be the total spin in the absolute ground state
of model (\ref{hamiltonian}) and let $S_{max}=(N+M)/2$.
Then in the low density limit
\be \lim_{|\L|\to\io}S/S_{max}\le \const \sqrt{\r^{1/3}a}\label{explicit}\ee
where the thermodynamic limit is taken keeping the total density $\r=
(N+M)/|\L|$ fixed.}\\

{\bf Remark.} It is natural to ask whether there exists some number $\r_c>0$
such that $\lim_{|\L|\to\io}S/S_{max}=0$ for all $\r<\r_c$ (see Problem 4 in 
[Li]). Note that Corollary 1 does not exclude this possibility. 
Note also that proving or disproving this possibility requires necessarily 
some non perturbative argument: any approximate computation of the ground 
state energy can only improve the error term in the r.h.s. of (\ref{explicit}) 
but will never establish the exact value of $S/S_{max}$. 

%%%%%%%%%%%%%%%%%%%%%%%%%%%%%%%%%%%%%%%%%%%%%%%%%%%%%%%%%%%%%%%%%%%%%%%%%%
%%%%%%%%%%%%%%%%%%%%%%%%%%%%%%%%%%%%%%%%%%%%%%%%%%%%%%%%%%%%%%%%%%%%%%%%%%
\section{The upper bound}
\setcounter{equation}{0}
%%%%%%%%%%%%%%%%%%%%%%%%%%%%%%%%%%%%%%%%%%%%%%%%%%%%%%%%%%%%%%%%%%%%%%%%%%
%%%%%%%%%%%%%%%%%%%%%%%%%%%%%%%%%%%%%%%%%%%%%%%%%%%%%%%%%%%%%%%%%%%%%%%%%%

In this section we shall assume $a/r_0> \d^{-1}(\r^{1/3}a)^{2/9}$, 
with $\d$ a constant to be chosen below. In this case it is enough to prove 
the upper bound
(\ref{0.3}) with $e_0(\r_{\uparrow},\r_{\downarrow})$ replaced by
$\frac 35 \left( 6\pi^2 \right)^{2/3}(\r_{\uparrow}^{5/3}+
\r_{\downarrow}^{5/3})$, see Remark (2) after the statement of the Theorem 
above. The weak coupling regime $a/r_0\le \d^{-1}
(\r^{1/3}a)^{2/9}$ is much simpler and will be treated in Appendix B.

It will be convenient to localize the particles
into small boxes with Dirichlet boundary conditions. The number of particles 
in each box will be large for small $\rho$, 
but finite and independent of the size of the large container $\L$. 
Let the side length of the small boxes be $\ell$. 
We then want to put $n = \r_{\uparrow} \ell^3$ spin-up particles into each box,
and likewise $m=\r_{\downarrow} \ell^3$ spin-down particles (here 
$\r_{\uparrow}=N|\L|^{-1}$
and $\r_{\downarrow}=M|\L|^{-1}$). Since $\rho_{\uparrow,\downarrow} 
\ell^3$ need not be an 
integer, however, we will choose 
\begin{equation}\label{defnm}
n=\r_{\uparrow} \ell^3 +\e_{\uparrow}\quad {\rm and} \quad m=\r_{\downarrow} 
\ell^3 +\e_\downarrow,\end{equation}
with $0\leq \e_{\uparrow,\downarrow} < 1$ chosen such that $n$ and $m$ are
integers. We then really have too many particles, but this is legitimate for 
an upper bound, since the energy is certainly increasing with particle number.
   
So, if $E_0(N,M,\L)$ is the ground state energy of (\ref{hamiltonian})
in the box $\O$, we have
\be\label{upper}
\lim_{|\L|\to\infty} \frac 1{|\L|} E_0(N,M,\L) \leq 
\frac{1}{\ell^3} E_0(n,m,\L_\ell),
\ee
where $\L_\ell$ is the cubic box of side $\ell$. Here we used that the 
interaction potential is zero range, so that different boxes of side $\ell$
are exactly decoupled. Note that actually the bound (\ref{upper}) is not 
only valid in the thermodynamic limit, but also for all finite cubic boxes 
$\L$ with side divisible by $\ell$.

We will now derive an upper bound on the ground state energy of $n$ 
spin-up and $m$ spin-down particles in a cubic box of side length $\ell$, 
for general $n$, $m$ and $\ell$. 
We take as a trial state the function 
\be\label{tri1}
\Psi(X,Y) = D_n(X) D_m(Y) G_n(X) G_m(Y) F(X,Y),
\ee
where $D_{n}(X)$ denotes the Slater determinant of the first $n$
eigenfunctions of the Laplacian in a  cubic box of side length $\ell$, with
Dirichlet boundary conditions; note that,
if $\phi_\a(x)$ are the eigenfunctions of the single-particle
Laplacian in a cubic box of side length $\ell$, we choose their normalization 
in such a way that $\sum_x r_0^3\phi_\a^*(x)\phi_\b(x)=\d_{\a,\b}$.
Moreover,
\be\label{tri2}
G_n(X)= \prod_{1\leq i<j\leq n} g(x_i-x_j),
\ee
with $0\leq g(x)\leq 1$, 
having the property that $g(x)=0$ for $|x|\leq s$ and $g(x)=1$ for $|x|
\geq 2s$, for some $s$ to be chosen 
later. We can assume that for any pair of nearest neighbor points $x$ and $x'$
we have $|g(x')-g(x)|\le 
\const r_0 s^{-1}$ for some constant independent of $s$. Finally,
\be\label{tri3}
F(X,Y) = \prod_{i=1}^n \prod_{j=1}^{m} f(x_i-y_j),
\ee
where, given a simply connected domain $\O\subset\ZZZ^3$ 
containing the origin, 
$f(x)=1$ if $x\not\in\O$. Inside $\O$ we choose $f(x)$ to be
the solution to the zero-energy 
scattering equation with boundary conditions $f(x)=\f(x)/\media{\f}_{\dpr\O}$
on the boundary $\dpr\O$ of the domain (here $\f(x)$ is given by 
(\ref{scattering}), $\dpr\O=\{x\in\O\,:\,
\dist(x,\O^c)=1\}$ and $\media{\f}_{\dpr\O}=
|\dpr\O|^{-1}\sum_{x\in\dpr\O}\f(x)$). We shall make the following explicit 
choice for the domain: $\O=B_R\cap\ZZZ^3$, where $B_R=\{x\in\RRR^3\,:\,
\f(x)\le 1-a/R\}$. Note that, if $R\gg r_0$, $B_R$ is approximately
a ball of radius $R$. Moreover $\media{\f}_{\dpr\O}=1-a/R+O(ar_0/R^2)$ and,
for any $x\in\dpr\O$, $\f(x)=\media{\f}_{\dpr\O}+O(ar_0/R^{2})$.
We assume $\d^{-1}r_0<
R\le s/5$, with $\d$ the same constant as in the condition
$a/r_0>\d^{-1}(\r^{1/3}a)^{2/9}$ (to be chosen below).

By the variational principle,
\be\label{vari}
E_0(n,m,\L_\ell)\leq \frac 
{\langle\Psi|H|\Psi\rangle}{\langle\Psi|\Psi\rangle},
\ee
with
$$\langle \Psi|H|\Psi\rangle =\langle\Psi| -\Delta_X|\Psi \rangle + 
\langle\Psi| -\Delta_Y|\Psi \rangle+U\langle\Psi|v_{XY}|\Psi \rangle.$$
(here, for any operator $\hat A$,
$\langle\Psi|\hat A|\Psi\rangle$ is defined as $\langle
\Psi|\hat A|\Psi\rangle=\sum_{X,Y}r_0^{3(n+m)}\Psi(X,Y)\,\hat A\,\Psi(X,Y)$
-- note the presence of the factor $r_0^{3(n+m)}$).
We first evaluate $\langle\Psi| -\Delta_X|\Psi \rangle$. By definition
it is equal to
\bea &&\frac1{r_0^2}\sum_{i=1}^{n}
\sum_{X,Y}r_0^{3(n+m)}D_m(Y)^2 G_m(Y)^2 D_n(X) G_n(X) F(X,Y)\cdot\\
&&\qquad\cdot
\sum_{x_i':|x_i'-x_i|=r_0}\Big[D_n(X)G_n(X)F(X,Y)-
D_n(X_i')G_n(X_i')F(X_i',Y)\Big]\nn\eea
where, if $X=\{x_1,\ldots,x_i,\ldots,x_{n_1}\}$, $X_i'$
is given by $X_i'=\{x_1,\ldots,x_i',\ldots,x_{n_1}\}$.
The r.h.s. of this equation can be written as
\bea&& \sum_{X,Y}r_0^{3(n+m)}D_m(Y)^2G_m(Y)^2 G_n(X)^2
F(X,Y)^2D_n(X) (-\D_X)D_n(X)\nn\\
&&+\frac1{r_0^2}\sum_{i=1}^{n}\sum_{X,Y}r_0^{3(n+m)}
\sum_{x_i':|x_i'-x_i|=r_0}D_m(Y)^2G_m(Y)^2D_n(X)D_n(X_i')\cdot\\
&&\hskip2.8truecm\cdot\, G_n(X)F(X,Y)
\Big[G_n(X)F(X,Y)-G_n(X_i')F(X_i',Y)\Big]\nn\eea
The first line is simply $E^D(n,\ell)\langle\Psi|\Psi\rangle$, 
where $E^D(n,\ell)$ is the sum of the lowest $n$ eigenvalues of the 
Dirichlet Laplacian in the box of side $\ell$ (note that these eigenvalues are
equal to $2r_0^{-2}\sum_{i=1}^3(1-\cos k_ir_0)$, with $k_i$ positive integer 
multiples of $\p/\ell$). An explicit computation shows that 
\be\label{kinetic} E^{\rm D}(n,\ell)\leq \frac 3 5 (6\pi^2)^{2/3}\frac{n^{5/3}}
{\ell^2}\left(1+\const n^{-1/3}+\const n^{2/3}(r_0/\ell)^2\right)\ee
The second line, if we 
symmetrize over $X,X_i'$, can be rewritten as
\bea\frac1{r_0^2}\sum_{i=1}^{n}\sum_{X\setminus x_i,Y}\sum_{<x_i,x_i'>}&&
r_0^{3(n+m)}D_m(Y)^2G_m(Y)^2D_n(X)D_n(X_i')\cdot\\
&&\cdot\,\Big[G_n(X)F(X,Y)-G_n(X_i')F(X_i',Y)\Big]^2\nn\eea
where $\sum_{<x_i,x_i'>}$ is the sum over the nearest neighbor bonds in 
$\L_\ell$. By Cauchy-Schwarz, we can bound the last expression from above
by
\bea\frac1{r_0^2}\sum_{i=1}^{n}\sum_{X\setminus x_i,Y}\sum_{<x_i,x_i'>}&&
r_0^{3(n+m)}D_m(Y)^2G_m(Y)^2D_n(X)^2\cdot\\
&&\cdot\Big[G_n(X)F(X,Y)-G_n(X_i')F(X_i',Y)\Big]^2\nn\eea
We now use the Schwarz inequality to deduce (for some $\e>0$ to be chosen 
later)
\bea&& \Big[G_n(X)F(X,Y)-G_n(X_i')F(X_i',Y)\Big]^2\nn\\
&&\le(1+\e)G_n(X_i')^2
\Big[F(X,Y)-F(X_i',Y)\Big]^2\\%\nn\\&&\hskip3.truecm
&&+(1+\e^{-1})
F(X,Y)^2\Big[G_n(X)-G_n(X_i')\Big]^2\nn\eea
Proceeding in the same way for the kinetic energy of the $Y$-particles, 
we thus get the upper bound
\be\label{vari2}
\langle\Psi|H|\Psi\rangle \leq \big[E^{\rm D}(n,\ell) + E^{\rm D}(m,\ell)\big] 
\langle\Psi|\Psi\rangle + (1+\e) I_2 + 
\left(1+\e^{-1}\right) I_3,\ee
with:
\bea\label{defII}
&&I_2=\sum_{X,Y}
r_0^{3(n+m)}D_n(X)^2D_m(Y)^2\,\cdot\nn\\
&&\hskip3.truecm\cdot\,
\Biggl\{G_m(Y)^2\frac12\sum_{i=1}^n\sum_{x'_i:|x_i'-x_i|=r_0}G_n(X_i')^2
\Big[\frac{F(X_i',Y)-F(X,Y)}{r_0}\Big]^2\nn\\
&&\hskip3.2truecm
+G_n(X)^2\frac12\sum_{j=1}^m\sum_{y'_j:|y_j'-y_j|=r_0}G_m(Y_j')^2
\Big[\frac{F(X,Y_j')-F(X,Y)}{r_0}\Big]^2\nn\\
&&\hskip3.2truecm
+UG_n(X)^2G_m(Y)^2v_{XY}F(X,Y)^2\Biggr\}\eea
and
\bea\label{defIII}
&&I_3=\sum_{X,Y}
r_0^{3(n+m)}D_m(Y)^2D_n(X)^2F(X,Y)^2\cdot\\
&&\hskip2.truecm\cdot
\Biggl\{G_m(Y)^2|\nabla_XG(X)|^2+G_n(X)^2|\nabla_YG(Y)|^2\Biggr\}\nn\eea
where $|\nabla_XG(X)|^2=\sum_{i=1}^n|\nabla_{i}G(X)|^2$ and
$|\nabla_iG(X)|^2=\frac12\sum_{\o=\pm}
|\nabla_{i}^\o G(X)|^2$. Moreover, for any 
function $g(x)$, the $\o$-gradient of $g$ is defined as
$\nabla^\o g(x)=\sum_{\ell=1}^3\o\hat e_\ell\big[g(x+\o r_0\hat e_\ell)-g(x)
\big]$ with $\hat e_\ell$ the coordinate versor in the $\ell$-th direction.
A similar definition is valid for $|\nabla_YG(Y)|^2$.
The positivity of $Uv_{XY}$ has been used here. 
Note that $E^{\rm D}(n,\ell)$ and $E^{\rm D}(m,\ell)$ can be bounded as in 
(\ref{kinetic}). We shall now bound $I_2$ and
$I_3$, when divided by $\langle\Psi|\Psi\rangle$, separately. 

Let us first derive an upper bound on $I_2$.
We are going to need the following lemma [LSS].\\

{\bf Lemma 1.} {\it
Let $D_n(X)$ denote a Slater determinant of $n$ linearly independent functions 
$\phi_\alpha(x)$. For a given
function $h(x)$ of one variable, let 
$\Phi(X)$ be the function $\Phi(X)=D_n(X)\prod_{i=1}^n h(x_i)$, 
and let $M$ denote the $n\times n$ matrix
\be\label{defM}
M_{\alpha\beta} = \sum_x r_0^3 
\phi_\alpha^*(x) \phi_\beta(x) |h(x)|^2\,.
\ee
Then the norm of $\Phi$ is given by $\langle\Phi|\Phi\rangle = \det M$.
Moreover, for $1\leq k\leq n$, the $k$-particle densities of $\Phi$ are given 
by
\bea&&
\pmatrix{n\cr k} \frac 1{\langle\Phi|\Phi\rangle} \sum_{x_{k+1},\ldots,x_n} 
r_0^{3(n-k)}|\Phi(X)|^2 =\\
&&\hskip1.truecm=
\frac 1{k!} \prod_{i=1}^k |h(x_i)|^2 \big(x_1\wedge \cdots \wedge x_k \big| 
M^{-1} \otimes \cdots \otimes M^{-1} \big|  x_1\wedge \cdots \wedge x_k 
\big),\nn\eea
where $|x)$ denotes the $n$-dimensional vector with components 
$\phi_\alpha(x)$, $1\leq \alpha\leq n$, and $|x_1 \wedge \cdots \wedge x_k)$ 
stands for the Slater determinant $(k!)^{-1/2} \sum_\sigma (-1)^{\sigma} 
|x_{\sigma(1)})\otimes \cdots\otimes |x_{\sigma(k)})$, $\sigma$ denoting 
permutations. 
Finally, if $\Phi'_i(X) = D_n(X) k(x_i)\prod_{j\neq i}h(x_j)$ 
for some function $k(x)$, then
\be\label{imme}
\sum_{i=1}^n \langle \Phi_i'|\Phi_i'\rangle = \Big(\det M \Big)\,
\Big( \Tr[K M^{-1}] \Big), 
\ee
where $\Tr[\, \cdot \, ]$ denotes the trace, and $K$ is the $n\times n$ matrix
\be\label{defK}
K_{\alpha\beta} = \sum_x r_0^3 \phi_\alpha^*(x) \phi_\beta(x) |k(x)|^2 \,.
\ee
}

Using $G_n(X_i')\leq 1$, we infer from this lemma that, for any fixed $Y$, 
\bea\nonumber
&&\sum_{X}r_0^{3n}D_n(X)^2\Biggl\{
\frac12\sum_{i=1}^n\sum_{x_i':|x_i'-x_i|=r_0}G_n(X_i')^2
\Big[\frac{F(X_i',Y)-F(X,Y)}{r_0}\Big]^2\\
&&\hskip7.truecm
+\frac{U}2G_n(X)^2v_{XY}F(X,Y)^2\Biggr\}\nn
\\
&&\le\sum_{X}r_0^{3n}D_n(X)^2\Biggl\{
|\nabla_XF(X,Y)|^2+\frac{U}2v_{XY}F(X,Y)^2\Biggr\}\label{eq13}\\
&&=\Tr\{K_Y M_Y^{-1}\}\sum_Xr_0^{3n}D_n(X)^2F(X,Y)^2\nn\eea
The $n\times n$ matrices $K_Y$ and $M_Y$ are given by (\ref{defM}) and 
(\ref{defK}), with $\phi_\alpha(x)$ being the lowest $n$ Dirichlet 
eigenfunctions of $-\Delta$, and with $h(x)=\prod_j f(x-y_j)$ and 
$$|k(x)|^2 = |\nabla h(x)|^2+ \frac{U}2
\sum_j\d_{x,y_j}\prod_jf(x-y_j)^2,$$
respectively (here $|\nabla h(x)|^2=(2 r_0^2)^{-1}\sum_{x': |x'-x|=r_0}|h(x)-
h(x')|^2$, see definition after (\ref{defIII})).

Since $K_Y$ is a positive definite matrix, we have the bound 
$\Tr K_Y M_Y^{-1}\leq \|M_Y^{-1}\|\Tr K_Y$, where $\|\, \cdot \, \|$ 
denotes the (spectral) matrix norm. 
To calculate $\Tr K_Y$, and to bound $\|M_Y^{-1}\|$, 
we can assume that all the $y_j$'s are separated by at least a distance $s$, 
because the summand in (\ref{defII}) vanishes otherwise. 

Since $s\geq 5R$ by assumption, we have in this case 
$|k(x)|^2= \sum_{j=1}^{n} \xi(x-y_j)$ with
\be\label{defxi}
\xi(x-y)= |\nabla f(x-y)|^2+\frac{U}2\d_{x,y}f(x-y)^2\,.\ee
Hence, if $\rho^{D}_n(x)$ denotes the one-particle density of $D_n(X)$, 
we have
\be\label{kytr}
\Tr K_Y = \sum_{j=1}^m \sum_x r_0^3\r^D_n(x)\x(x-y_j)\defin\sum_{j=1}^m
\rho^D_n* \xi(y_j),\ee
where $*$ denotes convolution. 
In order to bound $\|M_Y^{-1}\|$, we use the following:

{\bf Lemma 2.} {\it
Assume that $|y_i-y_j|\geq s\geq 5R $ for all $i\neq j$. Then
if $r_0/R$, $n^{-1}$ and $n (r_0/\ell)^3$ are sufficiently small
\be\| 1-M_Y \| \leq  \const \left(  
\frac {a R^2}{s^3} +  n^{2/3} \frac {s^2}{\ell^2}\right).\ee}

{\cs Proof.}
Let $q(x)= 1-\prod_j f(x-y_j)^2 \geq 0$. Then, 
for any $n$-dimensional vector $|b)$ with components $b_\alpha$, 
$$\big(b\big|1-M_Y\big|b\big) = \sum_x r_0^3  q(x) \Big |\sum_\alpha b_\alpha 
\phi_\alpha(x)\Big|^2 .$$
Hence, the question about the largest eigenvalue of $1-M_Y$
translates into the question of how large the average potential energy for the
potential $q(x)$ can be for functions such as $\sum_\alpha b_\alpha 
\phi_\alpha(x)$ whose kinetic energy is bounded above by $({\rm const.})\, 
n^{2/3} \ell^{-2}$, i.e., the Fermi energy for $n$ particles (under the 
assumption that $n\gg 1$ and $n r_0^3\ll \ell^3$).

Let $Q_j$ denote the cube of side $s/2$ centered at $y_j$. Note that all these 
cubes are non-overlapping by assumption. Also, since $s\geq 5R$, $q(x)=0$ if 
$x$ is outside all the cubes (recall that by definition -- 
see the lines following (\ref{tri3}) -- $f(x)$ is identically 1 outside a 
region $B_R$ of radius $R\left[1+O\big((r_0/R)^\k\big)\right]$, for some 
$\k>0$). 
For a given function $\phi(x)$, let $\phi_j$ 
denote the average of $\phi(x)$ in the cube $Q_j$. Moreover, let 
$\eta(x) = 
\phi(x) - \phi_j$. By the Cauchy-Schwarz inequality $(a+b)^2\leq 2(a^2+
b^2)$, we get the bound 
\be\sum_{x\in Q_j} q(x) |\phi(x)|^2  \leq 2 \sum_{x\in Q_j} q(x) 
|\eta(x)|^2 + 2 |\phi_j|^2 \sum_{x\in Q_j} q(x) .\ee
Note that $|\phi_j|^2 \leq 8s^{-3} \sum _{x\in Q_j}r_0^3 
|\phi(x)|^2$, again by the Cauchy-Schwarz inequality. 
Moreover, since $s\geq R$,
$$\sum_{x\in Q_j}r_0^3 q(x)=\sum_{x\in B_R}r_0^3(1-f(x)^2)\leq\const aR^2.$$
To obtain the last inequality, we used that if $x\in B_R$ then  
$f(x)=\f(x)/\media{\f}_{\dpr\O}\ge \f(x)$, with $\f(x)$ defined in 
(\ref{scattering}).

Note that $\eta(x)$ is a function whose average over the cube $Q_j$ 
is zero. In other words, it is orthogonal to the constant function
in $Q_j$. Hence, 
using the fact that $q(x)\leq 1$:
\bea\sum_{x\in Q_j}r_0^3 q(x) |\eta(x)|^2\, &&\leq \sum_{x\in Q_j} r_0^3 
|\eta(x)|^2\,\\&&\leq \frac1{2(1-\cos(2\p r_0s^{-1}))}
\sum_{x,x'\in Q_j\atop|x-x'|=r_0}r_0^3
|\eta(x)-\eta(x')|^2 \,,\nn\label{bea}\eea
where we used that $2r_0^{-2}(1-\cos(2\p r_0s^{-1}))$ 
is the second eigenvalue of the discrete Neumann Laplacian
in the cube of side $s/2$ and mesh $r_0$. 
In this last expression we can replace $\eta(x)$ 
by $\phi(x)$, of course, since they only differ by a constant. 
Summing over all the cubes $Q_j$ (and using that $q(x)=0$ outside the 
cubes), we thus obtain that, for any function $\phi(x)$,
$$\sum_{x}r_0^3  q(x) |\phi(x)|^2 \,\leq  
\const \left[ \frac {aR^2}{s^3}
\sum_x r_0^3 |\phi(x)|^2 \, + s^2 \sum_{x} r_0^3|\nabla\phi(x)|^2\,\right].$$
In the case in question, the kinetic energy of $\phi(x)$ is bounded by 
$\const n^{2/3}\ell^{-2}$. This finishes the proof of the lemma.
\qed

Since $0\leq M_Y\leq 1$ as a matrix, this lemma implies that 
\be\label{myesti}
\|M_Y^{-1}\| = \frac {1}{1-\|1-M_Y\|}\leq A_n \equiv \frac 1{ 1- \const 
\left[ aR^2 /s^3+ n^{2/3}(s/\ell)^2\right]}, \ee
provided the denominator is positive.
By inserting (\ref{kytr}) and (\ref{myesti}) into (\ref{eq13}), we see that, 
for fixed $Y$ with $|y_i-y_j|\geq s$ for all $i\neq j$,
\bea\nonumber
&&\sum_{X}r_0^{3n}D_n(X)^2\Biggl\{
|\nabla_XF(X,Y)|^2+\frac{U}2v_{XY}F(X,Y)^2\Biggr\}\\ 
&& \leq A_n \sum_{j=1}^n \rho^{\rm D}_n* \xi(y_j) \sum_X r_0^{3n}  
D_n(X)^2 F(X,Y)^2\,. \label{eq14}\eea
To be able later to compare this expression (\ref{eq14}) with $\langle 
\Psi|\Psi\rangle$, 
we want to put $G_n(X)^2$ back into the integrand. For this purpose we need 
the following lemma, which compares the integrals with and without the factor 
$G_n(X)^2$. \\

{\bf Lemma 3.} {\it For any fixed $Y$, if $n^{-1}$ and $n (r_0/\ell)^3$ 
are sufficiently small
\bea\nonumber
\sum_X &&r_0^{3n}  D_n(X)^2 F(X,Y)^2 G_n(X)^2\,\\ && \geq \sum_X r_0^{3n}  
D_n(X)^2 F(X,Y)^2 
\, \, \left(1- \const n^{8/3} \|M_Y^{-1}\|^2  (s/\ell)^{5}\right). 
\label{normg}\eea
}
{\cs Proof}
Since $g(x)=1$ for $|x|\geq 2s$, we have 
\be\label{seco}
G_n(X)^2 \geq 1 - \sum_{i<j}^n \theta(2s - |x_i-x_j|).\ee
Here $\theta$ denotes the Heaviside step function, i.e., $\theta(t)=1$ for 
$t\geq 0$ and $\theta(t)=0$ for $t<0$.
To evaluate the sum involving the second term in (\ref{seco}), we need the
two-particle density of the state $D_n(X)F(X,Y)$ for each fixed $Y$. By 
Lemma 1
above, and the fact that $f(x)\leq 1$, this density, when appropriately
normalized, is bounded from above by $\|M_Y^{-1}\|^2 \rho^{{\rm D},(2)}_n
(x,x')$, where $ \rho^{{\rm D},(2)}_n(x,x')$ denotes the 
two-particle density of the determinantal state $D_n(X)$. In particular, 
by explicit computation one finds that, if $n\gg1$ and $n r_0^3\ll \ell^3$,
this latter density satisfies the bound
\be\label{2pdd} 
\rho^{{\rm D},(2)}_n(x,x') \leq \const |x-x'|^2
(n/\ell^3)^{8/3}\ee
for some constant independent of $n$ and $\ell$.
Hence we arrive at (\ref{normg}).\qed

Let
$$B_n= \left(1- \const n^{8/3} A_n^2  (s/\ell)^{5}\right)^{-1},$$
assuming that the term in parenthesis is positive. 
Applying Lemma 3 to inequality (\ref{eq14}), we arrive at 
\bea\nonumber
&&\sum_{XY}r_0^{3(n+m)} 
G_n(X)^2 D_n(X)^2 \left[ |\nabla_X F(X,Y)|^2 + \frac{U}2 v_{XY} F(X,Y)^2
\right] D_m(Y) G_m(Y)\, \\ && \leq A_n B_n  \sum_{j=1}^n \sum_{XY}r_0^{3(n+m)} 
\rho^{\rm D}_n* \xi (y_j) D_m(Y)^2 D_n(X)^2 F(X,Y)^2 G_m(Y)^2 G_n(X)^2 \,. 
\label{eq15}\eea
Now we cannot bound $\rho^{\rm D}_n*\xi(y)$ independently of $y$ by simply 
using the supremum of $\rho^{\rm D}_n(x)$, since this number will be strictly 
bigger than $n/\ell^3$, even in the thermodynamic limit. Instead, we repeat 
the above argument for the $Y$ integration. We use $|G_m(Y)|\leq 1$, the 
$Y$-analogues of Lemma 1 and then Lemma 3 to put 
$G_m(Y)^2$ back in. Here, it is important to note that now the $x_i$'s 
are separated by at least a distance $s\geq 5R$. In this way we  obtain
\bea\nonumber
&&\sum_{X,Y}r_0^{3(n+m)} 
G_n(X)^2 D_n(X)^2 \left[ |\nabla_X F(X,Y)|^2 + \frac{U}2 v_{XY} F(X,Y)^2
\right] D_m(Y) G_m(Y)\,\\ && \leq  A_n B_n B_m  \sum_{XY}r_0^{3(n+m)}
D_m(Y)^2 
D_n(X)^2 F(X,Y)^2 G_m(Y)^2 G_n(X)^2 \, \Tr \widehat K_X M_X^{-1}
\label{ggg}\eea
The matrix $M_X$ is the same as before, with $Y$ replaced by $X$ (and $n$ 
replaced by $m$, of course), and $\widehat K_X$ is the $m\times m$ matrix
$$(\widehat K_X)_{\alpha\beta} = \sum_y r_0^{3}  
\phi_\alpha(y)^* \phi_\beta(y) 
\prod_{i}f(y-x_i)^2 \rho^{\rm D}_n*\xi(y)\,.$$
Using $|f(x)|\leq 1$ and $\|M_X^{-1}\|\leq A_m$, which follows from Lemma 2 
and the fact that the $x_i$'s are separated at least by a distance $s$, we get 
the bound
\be\label{last}
\Tr \widehat K_X M_X^{-1} \leq A_m \Tr \widehat K_X\leq A_m \sum_{x,y}r_0^6 
\rho^{\rm D}_n(x) \rho^{\rm D}_m(y) \xi(x-y) \,.\ee
A computation (see Appendix A) shows that 
\be\sum_x r_0^3\x(x)\le 4\pi a (1+\const a/R)\;.\label{ciao}\ee 
We then use this information to bound the 
last sum in (\ref{last}), by using Schwarz's inequality:
\bea&&\sum_{x,y}r_0^6 
\rho^{\rm D}_n(x) \rho^{\rm D}_m(y) \xi(x-y)\nn\\
&&\hskip1.truecm\le \left(
\sum_{x,y}r_0^6 \rho^{\rm D}_n(x)^2\xi(x-y)\right)^{1/2}
\left(\sum_{x,y}r_0^6 \rho^{\rm D}_m(y)^2\xi(x-y)\right)^{1/2}\nn\\
&&\hskip1.truecm=\left( \sum_x r_0^3 
\rho^{\rm D}_n(x)^2 \,
\right)^{1/2}\left( \sum_y r_0^3 \rho^{\rm D}_m(y)^2 \,\right)^{1/2} 
\sum_x r_0^3\x(x)\\
&&\hskip1.truecm\le \left( \sum_x r_0^3 
\rho^{\rm D}_n(x)^2 \,
\right)^{1/2}\left( \sum_y r_0^3 \rho^{\rm D}_m(y)^2 \,\right)^{1/2}   4\pi a 
\,(1+\const \frac{a}R).\nn\eea
For the square of $\rho^{\rm D}_n(x)$, by an explicit computation we find 
\be\label{inte3}
\sum_x r_0^3 \rho^{\rm D}_n(x)^2 \,\leq \frac {n^2}{\ell^3} \left (1+ 
\const n^{-1/3}+\const n^{2/3}(r_0/\ell)^2\right).\ee
The same holds with $n$ replaced by $m$. Eq. (\ref{ggg}) thus implies the 
upper bound
\bea
&&\sum_{XY}r_0^{3(n+m)}  
G_n(X)^2 D_n(X)^2 \left[ |\nabla_X F(X,Y)|^2 + \frac{U}2 v_{XY} F(X,Y)^2
\right] D_m(Y) G_m(Y)\,\nn\\ &&  \leq \langle\Psi|\Psi\rangle 
\frac{ 4\pi a n m}{\ell^3} A_n A_m B_n B_m\cdot\\
&&\hskip.2truecm\cdot  \left[1+ \const\left(\frac{a}R 
+n^{-1/3} + m^{-1/3}+ (n+m)^{2/3}(r_0/\ell)^2\right)\right].\nn
\label{estiIIa}\eea
The same bound holds, of course, with $X$ and $Y$ interchanged. We therefore 
have the upper bound
\bea&& I_2  
\leq \langle\Psi|\Psi\rangle \frac{ 8\pi a n m}{\ell^3} A_n A_m B_n B_m
\cdot\label{estiII}\\  
&&\hskip.8truecm\cdot\left[1+ \const\left(\frac{a}R 
+n^{-1/3} + m^{-1/3}+ (n+m)^{2/3}(r_0/\ell)^2\right)
\right].\nn
\eea

It remains to bound the term $I_3$. Using $|g(x)|\leq 1$ we have that  
\bea\nonumber
|\nabla_X G_n(X)|^2&\leq& \frac12\sum_{\o=\pm}\Biggl[ 
\sum_{i=1}^n \ \sum_{j,\, j\neq i}
|\nabla^\o g(x_i-x_j)|^2  \\ && + \sum_{i=1}^n \ \sum_{j,\, j\neq i} \ 
\sum_{k,\, k\neq i,j}
 |\nabla^\o g(x_i-x_j)\cdot\nabla^\o g(x_i-x_k)|\Biggr],\label{susu}\eea
where the $\o$-gradient $\nabla^\o$ was defined after (\ref{defIII}).
Now, by Lemma 1, the appropriately normalized $k$-particle
densities of $D_n(X)F(X,Y)$ are bounded above by $\|M_Y^{-1}\|^k
\rho^{{\rm D},(k)}_n$, where $ \rho^{{\rm D},(k)}_n$ denotes the
$k$-particle density of $D_n(X)$. In particular, $\rho^{{\rm
D},(2)}_n$ satisfies the bound (\ref{2pdd}), and $\rho^{{\rm D},(3)}_n$
satisfies
$$\rho^{{\rm D},(3)}_n(x,x',x'')\leq \const (n/\ell^3)^3$$
for some constant independent of $n$ and $\ell$. 
Remember that, by the definition of $g$ (see lines following (\ref{tri2})),
if $x$ and $x'$ are two neighbor points, 
$g(x')-g(x)$ is zero for $|x|>2s$ and otherwise 
$|g(x')-g(x)|\leq \const r_0 s^{-1}$. Using this, 
we obtain from (\ref{susu}), for any fixed $Y$,
\bea
&& \sum_{X} r_0^{3n} 
D_n(X)^2 F(X,Y)^2 |\nabla_X G_n(X)|^2\,\label{sec}\\ &&\leq \const 
\frac {n^2}{\ell^3} s \left( \|M_Y^{-1}\|^{2} n^{2/3} (s/\ell)^{2} + 
\|M_Y^{-1}\|^3 n (s/\ell)^3 \right) \sum_X r_0^{3n} D_n(X)^2 F(X,Y)^2\,. 
\nn\eea
Finally, to get a bound on $I_3$, we proceed as above, using (\ref{myesti}) 
(and the fact that the $y_j$'s are separated by a distance $s$) and Lemma 3 
to put $G_n(X)^2$ back into the integral. Note, however, that it is enough 
to bound $A_n$ and $B_n$ by constants in this term. Assuming that 
$n(s/\ell)^3$ is small, the second term in the parenthesis in (\ref{sec}) 
is negligible compared to the first term. The same bound applies to the 
case where $X$ and $Y$ are interchanged, and hence we obtain
\be\label{estiIII}
I_3 \leq \langle\Psi|\Psi\rangle\, \const \big(n^{8/3} +m^{8/3}\big)
\frac {s^3}{\ell^5} .\ee

Collecting all the error terms obtained in Eqs. (\ref{kinetic}), 
(\ref{estiII}) 
and (\ref{estiIII}) and inserting them into (\ref{vari}) and (\ref{vari2}), 
we obtain 

\bea\nonumber
&& E_0(n,m,\L_\ell)\leq \\
&&\hskip1.truecm\leq \frac 3 5  (6\pi^2)^{2/3} \frac{n^{5/3}+m^{5/3}}{\ell^2}
\left(1+C n^{-1/3}+C m^{-1/3}+C (n+m)^{2/3}(r_0/\ell)^2\right) 
\nn\\ \nonumber &&\hskip1.truecm+ 8\pi a\frac{nm}{\ell^3}
\left( 1 + \e+  C \left[ \frac {a R^2}{s^3}+ (n+m)^{2/3} (s/\ell)^2 \right.
\right.\\
&&\hskip1.truecm\left.\left.
+\frac aR +\frac 1{n^{1/3}}+\frac 1{m^{1/3}} + (n+m)^{8/3} (s/\ell)^5 \right]
\right)\label{ineq33}\\  &&\hskip1.truecm
+ \frac {Cs}\e\frac { (n+m)^2}{\ell^3} \left[ (n+m)^{2/3} 
(s/\ell)^2\right]\nn\eea
for some constant $C>0$. In Ineq. (\ref{ineq33}) we have assumed smallness of 
all the error terms, i.e., that the terms in square brackets are small. This 
condition will be fulfilled, at low density, with our choice of $R$, $s$, 
$n$, $m$ and $\ell$ below. 

The optimal choice of $\e$ in (\ref{ineq33}) is given by $\e^2 = \const 
(n+m)^{8/3} s^3/(\ell^2 a n m)$. Inserting this value for $\e$ we infer 
from (\ref{ineq33})
\bea\nonumber
&& E_0(n,m,\L_\ell) \leq\\
&&\hskip1.truecm\leq\nn
 \frac 3 5  (6\pi^2)^{2/3} \frac{n^{5/3}+m^{5/3}}{\ell^2}
\left(1+C n^{-1/3}+C m^{-1/3}+C (n+m)^{2/3}(r_0/\ell)^2\right) 
\\ \nonumber &&\hskip1.truecm+ 8\pi a\frac {nm}{\ell^3}
\left( 1 +  C \left[ \frac {a R^2}{s^3}+ (n+m)^{2/3} (s/\ell)^2\right.\right.\\
&&\hskip1.truecm\left.\left. +\frac aR +
\frac 1{n^{1/3}}+\frac 1{m^{1/3}} + (n+m)^{8/3} (s/\ell)^5 \right]\right)\\  
&&\hskip1.truecm+ C(n+m)^{7/3} \frac {s^{3/2} a^{1/2}}{\ell^4} .\nn  
\label{ineq33s}\eea
Eq. (\ref{ineq33s}) is our final bound on the energy $E_0(n,m,\ell)$. 
To apply this result in (\ref{upper}) we have to insert the values
(\ref{defnm}) for $n$ and $m$.
Recall that $|n-\r_{\uparrow}\ell^3|\le 1$ and $|m-\r_{\downarrow}\ell^3|\le1$.
We are then still free to choose $R$, $s$ and $\ell$. We choose
$$R=a \big(a\rho^{1/3}\big)^{-2/9}\ , \ s=6 R \ , \ 
 \ell = \rho^{-1/3} \big(a\rho^{1/3}\big)^{-11/9}.$$
Note that with these choices the condition $a/r_0>\d^{-1}(\r^{1/3}a)^{2/9}$
implies $r_0/R<\d$. We choose $\d$ to be so small that Lemma 2 is valid.
Inserting these values into (\ref{ineq33s}) we thus obtain, for small $\rho$, 
$$\frac 1{\ell^3} E_0(n,m,\L_\ell) \leq 
\frac 3 5  (6\pi^2)^{2/3} \big[\r_{\uparrow}^{5/3}+ 
\r_{\downarrow}^{5/3}\big]+ 8\pi a 
\r_{\uparrow}\r_{\downarrow}+ \const  a \rho^2  \big(a\rho^{1/3}\big)^{2/9}.$$
In combination with Eq. (\ref{upper}), this finishes the proof of the upper 
bound in the case $a/r_0\ge \d^{-1}(\r^{1/3}a)^{2/9}$. The opposite case
(that is much simpler) is treated in Appendix B.

\acknowledgments I would like to thank E. H. Lieb and R. Seiringer for 
several helpful discussions and comments. This work was partially supported 
by U.S. National Science Foundation grant PHY 01 39984.

%%%%%%%%%%%%%%%%%%%%%%%%%%%%%%%%%%%%%%%%%%%%%%%%%%%%%%%%%%%%%%%%%%%%%%%%%%%%
\appendix
\section{Proof of (\ref{ciao})}\lb{A1}
\setcounter{equation}{0}
\renewcommand{\theequation}{\ref{A1}.\arabic{equation}}
%%%%%%%%%%%%%%%%%%%%%%%%%%%%%%%%%%%%%%%%%%%%%%%%%%%%%%%%%%%%%%%%%%%%%%%%%%%%
%%%%%%%%%%%%%%%%%%%%%%%%%%%%%%%%%%%%%%%%%%%%%%%%%%%%%%%%%%%%%%%%%%%%%%%%%%%%

In this Appendix we want to show that if $\x(x)$ is defined by (\ref{defxi})
then $\sum_xr_0^3\x(x)\le 4\p a(1+\const a/R)$. Note that ``integrating by 
parts'' we can rewrite the summation as:
\be\sum_xr_0^3\x(x)=\sum_{x}r_0^3f(x)(-\D_x)f(x)+\frac{U}2f(0)^2\label{A.1}\ee
Note also that, by definition, if $x\in\O$ then $f(x)$ coincides with $\f(x)/
\media{\f}_{\dpr\O}$, while if $x\not\in\O$ then $f(x)=1$. Then, by
the definition of $\f(x)$, we see that in the summation in (A.1)
all terms with $x$ ``well inside'' $\O$ (\ie with $x$ such that
$\dist(x,\O^c)\ge 2r_0$) cancel out with $\frac{U}2f(0)^2$, and all terms
with $x$ ``well outside'' $\O$ (\ie with $x$ such that $\dist(x,\O)\ge 2r_0$))
are identically zero. 
So we are left with a boundary term, that is a summation over the $x$'s at a 
distance $r_0$ from $\O$ or from $\O^c$. Let us recall that $\dpr\O=\{x\in\O\,:
\,\dist(x,\O^c)=r_0\}$ and let us define $\dpr\O^c=\{x\in\O^c\,:\,
\dist(x,\O)=r_0\}$. The r.h.s. of (A.1) can be rewritten as
\bea&& \sum_{x\in\dpr\O}r_0^3
f(x)(-\D)f(x)+\sum_{x'\in\dpr\O^c}r_0^3f(x')(-\D)f(x')\\
&&=\sum_{x\in\dpr\O}r_0
\frac{\f(x)}{\media{\f}_{\dpr\O}}\sum_{x'\in\dpr\O^c}^{(x)}
\left[\frac{\f(x')}{\media{\f}_{\dpr\O}}-1\right]+
\sum_{x'\in\dpr\O^c}r_0
\sum_{x\in\dpr\O}^{(x')}\left[1-\frac{\f(x)}{\media{\f}_{
\dpr\O}}\right]
\label{A.2}\nn\eea
where $\sum_{x'\in\dpr\O^c}^{(x)}$ is the sum over the points $x'\in\dpr\O^c$
at a distance $r_0$ from $x$ (and similarly for $\sum_{x\in\dpr\O}^{(x')}$).
The second line in (A.2) can still be rewritten as
\be \sum_{x\in\dpr\O}r_0\left[\frac{\f(x)}{\media{\f}_{\dpr\O}}-1\right]
\sum_{x'\in\dpr\O^c}^{(x)}
\left[\frac{\f(x')}{\media{\f}_{\dpr\O}}-1\right]+
\frac1{\media{\f}_{\dpr\O}}
\sum_{<x,x'>}^*r_0\left[\f(x')-\f(x)\right]\label{A.3}\ee
where $\sum_{<x,x'>}^*$ is the sum over the nearest neighbor pairs $<x,x'>$
with $x\in\dpr\O$ and $x'\in\dpr\O^c$.
Recall that, if $R\gg r_0$, then 
for any $x\in\dpr\O$ and any $x'\in\dpr\O^c$ we have
$\f(x),\f(x')=\media{\f}_{\dpr\O}+O(ar_0/R^2)$. Then 
the first term in (A.3) can be bounded above by a constant times 
$(R^2/r_0^2)r_0(ar_0/R^2)^2=a(ar_0/R^2)< a(a/R)$. Moreover, note that
the second term in (A.3) is proportional to the (discrete) flux of 
the discrete derivative of $\f$ 
across the ``surface'' of $\O$ and the latter is {\it equal} to $4\p a$,
see (\ref{0.2}). As a conclusion, the second term
in (A.3) is equal to $4\p a/\media{\f}_{\dpr\O}$.
Using that $\media{\f}_{\dpr\O}=1-a/R+O(ar_0/R^2)$, see lines following 
Eq. (\ref{tri3}), (\ref{ciao}) is proven.

%%%%%%%%%%%%%%%%%%%%%%%%%%%%%%%%%%%%%%%%%%%%%%%%%%%%%%%%%%%%%%%%%%%%%%%%%%%%
\section{The weak coupling regime}\lb{A2}
\setcounter{equation}{0}
\renewcommand{\theequation}{\ref{A2}.\arabic{equation}}
%%%%%%%%%%%%%%%%%%%%%%%%%%%%%%%%%%%%%%%%%%%%%%%%%%%%%%%%%%%%%%%%%%%%%%%%%%%%
%%%%%%%%%%%%%%%%%%%%%%%%%%%%%%%%%%%%%%%%%%%%%%%%%%%%%%%%%%%%%%%%%%%%%%%%%%%%

In this Appendix we prove the main Theorem in the case that 
$a/r_0\le \d^{-1}(\r^{1/3}a)^{2/9}$. In this case we do not localize particles
and we simply choose as trial function the ground state of the free Fermi gas:
$\Psi(X,Y)=D_N(X)D_M(Y)$, where $D_N(X)$ denotes 
the Slater determinant of the first $N$
eigenfunctions of the Laplacian in the cubic box $\L$ 
with (say) periodic boundary conditions (a similar definition is valid for 
$D_M(Y)$). We assume that $D_N(X)$ and $D_M(Y)$ are normalized in
such a way that $\media{\Psi|\Psi}=\sum_{X,Y}r_0^{3(N+M)}|D_N(X)|^2|D_M(Y)|^2
=1$. By the variational principle $E_0(N,M,\L)\leq\langle\Psi|H|\Psi\rangle
=E_0^{(U=0)}(N,M,\L)+U\sum_{i=1}^N\sum_{j=1^M}\langle\Psi|v_{X,Y}
|\Psi\rangle$. Since the specific energy corresponding to the term 
$E_0(N,M,\L)|_{U=0}$ is by definition $e_0(\r_{\uparrow},\r_{\downarrow})$,
we are left with bounding 
\be \sum_{Y}r_0^{3M}|D_M(Y)|^2\sum_{i=1}^N\sum_Xr_0^{3N}|D_N(X)|^2\sum_{j=1}^M
\d_{x_i,y_j}\ee
An application of Lemma 1 shows that this expression is equal to 
$\sum_x r_0^6 \r_N(x)\r_M(x)$, where $\r_N$ is the $1$-particle density of
$D_N(X)$ and $\r_M$ is the $1$-particle density of $D_N(Y)$. In our case
$\r_{N}(x)\=\r_{\uparrow}$ and $\r_M(x)\=\r_{\downarrow}$. Then we get
$U\sum_{i=1}^N\sum_{j=1^M}\langle\Psi|v_{X,Y}
|\Psi\rangle=|\L|Ur_0^3\r_{\uparrow}\r_{\downarrow}$. 
By (\ref{length}) we have that $Ur_0^3=8\p a(1+\g Ur_0^2)$. As a conclusion:
\be 
\lim_{|\L|\to\io}\frac1{|\L|}E_0(N,M,\L)\le e_0(\r_{\uparrow},\r_{\downarrow})
+8\p a\r_{\uparrow}\r_{\downarrow}(1+\g Ur_0^2)\ee
Since $Ur_0^2=8\p a/r_0(1+\const a/r_0)$, we have that $\g Ur_0^2\le
\const (\r^{1/3}a)^{2/9}$ and the proof is concluded.\\

{\centerline{\bf References}}
\\
\\
[BLT] V. Bach, E. H. Lieb, M. V. Travaglia: {\it Ferromagnetism of the 
Hubbard model at strong coupling in the Hartree--Fock approximation}, Rev. 
Math. Phys. {\bf 18}, 519--543 (2006).
\\
[D] F. J. Dyson: {\it Ground-State Energy of a Hard-Sphere Gas}, Phys. Rev. 
{\bf 106}, 20--26 (1957).
\\
[Le] W. Lenz: {\it Die Wellenfunktion und Geschwindig\-keits\-verteilung des 
entarteten Gases}, Z. Phys. {\bf 56}, 778--789 (1929).\\
[Li] E. H. Lieb: {\it The Hubbard model: some rigorous results and open 
problems}, Advances in dynamical systems and quantum physics (Capri, 1993),  
173--193, World Sci. Publ., River Edge, NJ, 1995.
\\
[LSS] E. H. Lieb, R. Seiringer, and J. P. Solovej:
{\it Ground-state energy of the low-density Fermi gas},
Phys. Rev. A {\bf 71}, 053605 (2005).\\
[LSSY] E. H. Lieb, R. Seiringer, J. P. Solovej, J. Yngvason:{\it 
The mathematics of the Bose gas and its condensation}, Oberwolfach Seminars, 
{\bf 34}. Birkh\"auser Verlag, Basel, 2005.
\\
[LY] E. H. Lieb, J. Yngvason: {\it Ground State Energy of the Low Density Bose 
Gas}, Phys. Rev. Lett. {\bf 80}, 2504-2507 (1998).
\\
[SR] L. Spruch, L. Rosenberg: {\it Upper bounds on scattering lengths for 
static potentials}, Phys. Rev. {\bf 116}, 1034 (1959).
\\
[T] H. Tasaki: {\it The Hubbard model -- an introduction and selected rigorous
results}, J. Phys.: Condens. Matter {\bf 10}, 4353-4378 (1998).

\end{document}